\documentclass[useAMS,usenatbib]{mn2e}
\usepackage{amsmath}
\usepackage{graphicx}
\usepackage{txfonts}
\usepackage{natbib}
\usepackage{bm}

\def\beq#1{\begin{equation}\label{#1}}
\def\eeq{\end{equation}}
\def\beqa#1{\begin{eqnarray}\label{#1}}
\def\eeqa{\end{eqnarray}}

\def\Eq#1{Eq.~(\ref{#1})} 
\def\eqn#1{~(\ref{#1})}
\def\myfrac#1#2{\left(\frac{#1}{#2}\right)}

\def\comment#1{\relax}

\title[Viscous instability in axisymmetric flows]{A viscous instability in axially 
symmetric laminar shear flows}
\author[N. Shakura \& K. Postnov] {N. Shakura
\thanks{E-mail: nikolai.shakura@gmail.com, kpostnov@gmail.com},
K. Postnov\\
Sternberg Astronomical Institute, Moscow M.V. Lomonosov State University, Universitetskij pr., 13, 119992, Moscow, Russia}	

\begin{document}

\date{Received ... Accepted ...}
\pagerange{\pageref{firstpage}--\pageref{lastpage}} \pubyear{2012}

\maketitle

\label{firstpage}

\begin{abstract}
A viscous instability in shearing laminar axisymmetric
hydrodynamic flows around a gravitating center is described.
In the linearized hydrodynamic equations written in the Boussinesq approximation
with microscopic molecular transport coefficients, 
the instability arises when the viscous dissipation is taken into account in 
the energy equation. Using the local WKB approximation, we derive a third-order algebraic dispersion equation with two
modes representing the modified Rayleigh modes R+ and R-, and the third X-mode.
We show that in thin accretion flows the viscosity destabilizes 
one of the Rayleigh modes in a wide range of wavenumbers, while the X-mode always remains stable.  
In Keplerian flows, the instability increment is found to be a few Keplerian
rotational periods at wavelengths with $kr\sim 10-50$.
This instability may cause turbulence in astrophysical 
accretion discs even in the absence of magnetic field.

\end{abstract}


\section{Introduction}
\label{intro}

The origin of turbulence in accretion discs is an outstanding problem in 
astrophysics. The dimensionless phenomenological parameter $\alpha$ introduced by \cite{1973A&A....24..337S} for assumed turbulent eddy viscosity 
and chaotic magnetic fields 
turned out to be very useful in describing physical properties of accretion discs. 
Analysis of 
different observations (e.g., the behaviour of non-stationary accretion discs in X-ray novae 
\citep{2008A&A...491..267S} and dwarf-nova and AM CVn stars 
\citep{2012A&A...545A.115K}) suggest a rather 
large values $\alpha\sim 0.3$, indicating the presence of well-developed turbulence
in the disc. In Keplerian accretion discs, the angular momentum increases with radius,
making the flow stable against small hydrodynamic perturbations
according to the classical Rayleigh criterion. 
When the small magnetic field is present in fully ionized gas, 
a popular mechanism quenching the instability is the Velikhov-Chandrasekhar 
magneto-rotational instability (MRI) \citep{Velikhov59, 1960PNAS...46..253C, 1991ApJ...376..214B} 
(see \cite{1998RvMP...70....1B} for a detailed review). In spite of being a powerful
instability, MRI has its own limitations (see e.g. \cite{1994ApJ...432..213G}, 
for discussion of parasiting instabilities 
and \cite{2014arXiv1412.1223S}, for discussion of 
applications to thin Keplerian accretion discs). 

As for the purely hydrodynamic case, 
so far there has been no clear criterion of the hydrodynamic turbulence. 
In a Keplerian flow, there are different mechanisms for small perturbations growth, such as 
linear growth of transient perturbations (see the recent study \cite{2014MNRAS.442..870Z}
and references therein), but the transition of these perturbations 
to turbulence, which is a strongly non-linear process, remains unclear.

In this paper we perform the linear stability analysis 
of shearing laminar hydrodynamic viscous flows with arbitrary rotation laws in the form 
$\Omega^2\propto r^{-n}$
taking into account the viscous heating and thermal conductivity 
in the energy equation. 
Unlike many previous works, 
we use microscopic molecular transport coefficients to describe the viscosity and heat conductivity.
These terms in the
energy equation make one of the Rayleigh modes unstable (i.e. make 
their amplitude exponentially 
growing) in a wide range of wave numbers for perturbations 
normal to the direction of the wave vector. 
Physically, the instability 
may be due to
the viscously heated gas being unstable to convection in the gravity field of the 
central object in the absence of the background entropy gradients.

The instability increment 
decreases (but does not vanish) with increasing thermal conductivity
and is maximum in cold neutral flows with largest Prandtl numbers. 
We discuss the relevance of the found 
viscous instability to the generation of turbulence in 
laminar thin accretion flows.

The pulsational instability of viscous accretion discs was first 
studied by \cite{1978MNRAS.185..629K}.
The viscous instability of the standard turbulent 
Shakura-Sunyaev $\alpha$-discs 
was investigated in many papers (see, e.g., 
\cite{1984ApJ...287..774B, 1993ApJ...409..739K, 2006MNRAS.372.1829L}, among others). (Note that in the latter papers the viscous instability is
referred to as 'viscous overstability', i.e. when arising as an
exponentially growing oscillations, in analogy with stellar pulsations 
discussed by \cite{1926ics..book.....E}, and the term 'instability' is reserved for purely imaginary negative modes.)

In Section \ref{s:deriv} we derive the basic dispersion equation. To make 
the physical case as simple as possible, we work in the Boussinesq approximation 
(i.e. consider the fluid incompressible $\nabla \bm{u}=0$,
keep the Eulerian pressure variations non-zero only in the equations of motion
and put them zero in the energy equation) and take small perturbations in the form 
of plane waves in the direction transversal to the wave propagation. 
For such perturbations we derive a third-order algebaraic dispersion 
equation with imaginary coefficients, which has three solutions: one
Rayleigh mode with positive real part (R+), the Rayleigh mode with negative real 
part (R-) and a new mode which has zero real and imaginary part at $k\to 0$ (the X-mode).
We find that one of the Rayleigh modes becomes unstable (exponentially growing) at large
wavelengths, while the X-mode remains stable at all wavelengths.  
In Section \ref{s:analysis} we analyze the obtained the dispersion equation. First we rewrite 
it in the dimensionless form, and then perform its numerical analysis for several important cases 
of thermal conductivity (for purely electron conductivity in the fully ionized plasma,
for the case where the radiation conductivity is important, and for the case of neutral
monoatomic hydrogen gas where the heat conductivity and viscosity are caused by 
the same particles). 
In Section \ref{s:disc} we discuss the applicability 
of the approximation of incompressibility we use and damping of the instability 
due to entropy gradients in the unperturbed flow. 
Section \ref{s:concl} summarizes our findings.   
Details of linearization of the viscous force in the dynamical equations
are given in the Appendix.

\section{Derivation of the dispersion equation}
\label{s:deriv}

We start with  the linear analysis of hydrodynamic equations. The fluid viscosity 
and thermal conductivity is taken into account through kinematic viscosity coefficient $\nu$ and
heat conductivity coefficient $\kappa$, respectively.

\subsection{Basic equations}

The system of hydrodynamic equations reads:

\begin{enumerate}
\item{mass conservation equation}
\beq{e:mass}
\frac{\partial \rho}{\partial t}+\nabla\cdot(\rho\bm{u})=0\,,
\eeq
In cylindrical coordinates for axially symmetric flows:
\beq{e:div}
\nabla\cdot(\rho\bm{u})=\frac{1}{r}\frac{\partial (\rho r u_r) }{\partial r}
+\frac{\partial (\rho u_z) }{\partial z}
\eeq

\item{Navier-Stokes equation including gravity force} 
\beq{e:NS}
\frac{\partial \bm{u}}{\partial t}+(\bm{u}\nabla)\cdot\bm{u}=-\frac{1}{\rho}\nabla p -
\nabla \phi_g +\bm{{\cal N}}\,.
\eeq
Here $\phi_g=-GM/r$ is the Newtonian gravitational potential of the central body with mass $M$, $\bm{{\cal N}}$ is the viscous force. In cylindrical coordinates for axisymmetric flows:
\beq{e:NSr}
\frac{\partial u_r}{\partial t}+ u_r \frac{\partial u_r}{\partial r}+
u_z\frac{\partial u_r}{\partial z}-\frac{u_\phi^2}{r}=
-\frac{\partial \phi_g}{\partial r}-\frac{1}{\rho}\frac{\partial p}{\partial r}+{\cal N}_r\,,
\eeq

\beq{e:NSphi}
\frac{\partial u_\phi}{\partial t}+ u_r \frac{\partial u_\phi}{\partial r}+
u_z\frac{\partial u_\phi}{\partial z}+\frac{u_r u_\phi}{r}={\cal N}_\phi\,,
\eeq

\beq{e:NSz}
\frac{\partial u_z}{\partial t}+ u_r \frac{\partial u_z}{\partial r}+
u_z\frac{\partial u_z}{\partial z}=
-\frac{\partial \phi_g}{\partial z}-\frac{1}{\rho}\frac{\partial p}{\partial z}+{\cal N}_z\,.
\eeq
The linearized viscous force components are specified in Appendix A. 

\item{energy equation} 
\beq{e:en}
\frac{\rho {\cal R} T}{\mu}\left[
\frac{\partial s}{\partial t}+(\bm{u}\nabla)\cdot s
\right]=
Q_\mathrm{visc}-\nabla\cdot\bm{F}\,.
\eeq
where $s$ is the specific entropy per particle, 
$Q_\mathrm{visc}$ is the viscous dissipation rate per unit volume, 
${\cal R}$ is the universal gas constant, 
$\mu$ is the molecular weight, $T$ is the temperature, and terms on the right 
stand for the viscous energy production and the heat conductivity energy flux $\bm{F}$, 
respectively. The energy flux due to the heat conductivity is 
\beq{e:kappat}
\nabla\cdot\bm{F}=\nabla(-\kappa\nabla T)=-\kappa\Delta T-\nabla\kappa\cdot\nabla T\,.
\eeq
Note that both electrons and photons, and
at low temperatures neutral atoms, can contribute to the heat conductivity
(see Section \ref{s:analysis} below).

\item{equation of state}

The equation of state for a perfect gas is convenient to write in the form:
\beq{e:eos}
p=Ke^{s/c_V}\rho^\gamma\,,
\eeq 
where $K$ is a constant, $c_V=1/(\gamma-1)$ is the specific volume heat capacity 
and $\gamma=c_p/c_V$ is the adiabatic index (5/3 for the monoatomic gas). 

\end{enumerate}

\subsection{Linearization of basic equations}

We will consider small axially symmetric perturbations in 
the WKB approximation with space-time dependence $e^{i(\omega t-k_r r-k_z z)}$, where $r,z,\phi$ are cylindrical coordinates. 
The velocity perturbations are $\bm{u}=(u_r,u_\phi,u_z)$. 
The density, pressure, temperature and entropy perturbations 
are $\rho_1$, $p_1$, $T_1$, and $s_1$ over the unperturbed values 
$\rho_0$, $p_0$, $T_0$, and $s_0$, respectively. As a simplification, 
to filter out 
acoustic oscillations arising from the restoring pressure force, we
will use the Boussinesq approximation, i.e. consider incompressible gas motion 
$\nabla\cdot \bm{u}=0$. 
In the energy equation we will
neglect Eulerian pressure variations,
$p_1(t, r, \phi, z) =0$  (see the justification below), but Lagrangian pressure variations
$\delta p(t, r(t_0), \phi(t_0, z(t_0))$ are non-zero. (We remind that for infinitesimally 
small shifts a perturbed gas parcel acquires the pressure equal to that of the ambient medium; 
see e.g. \cite{1960ApJ...131..442S,  Kundu5ed} for discussion of the Boussinesq approximation). 
We stress that 
we investigate the motion of axisymmetric transverse perturbations, i.e. 
small perturbations in the direction normal to the wave vector $\bm{k}$.    

\subsubsection{Dynamical equations}

In the linear approximation, the system of differential hydrodynamic 
equations is reduced to the following system of algebraic equations.

a).  
The Boussinesq approximation for gas velocity $\bm{u}$ is $\nabla\cdot  \bm{u}=0$:
\beq{Bq}
k_r u_r+k_zu_z=0\,.
\eeq

b). 
The radial, azimuthal and vertical components of the Navier-Stokes momentum equation are, respectively:
\beq{iur}
i\omega u_r-2\Omega u_\phi=ik_r\frac{p_1}{\rho_0}-\frac{\rho_1}{\rho_0^2}\frac{\partial
p_0}{\partial r}-\nu k^2u_r[R]\,,
\eeq
where the factor $[R]$
takes into account the dependence of the viscosity coefficient 
on temperature $\eta\sim T^{\alpha_{visc}}$ ($\alpha_{visc}=5/2$ for fully ionized gas and $\alpha_{visc}=1/2$ for neutral gas) in the perturbed viscous force component ${\cal N}_r$ (see \Eq{e:[R]} in Appendix A);

\beq{iuphi}
i\omega u_\phi+\frac{\varkappa^2}{2\Omega}u_r=-\nu k^2u_\phi[\Phi]\,,
\eeq
where the factor $[\Phi]$
takes into account variations of the viscosity coefficient in the perturbed viscous force component ${\cal N}_\phi$ (see \Eq{e:[Phi]} in Appendix A);

\beq{iuz}
i\omega u_z=ik_z\frac{p_1}{\rho_0}-\frac{\rho_1}{\rho_0^2}\frac{\partial p_0}{\partial z}-\nu k^2u_z[Z]
\eeq
where the factor $[Z]$
takes into account variations of the viscosity coefficient in the perturbed viscous force component ${\cal N}_z$ (see \Eq{e:[E]} in Appendix A).
Here $k^2=k_r^2+k_z^2$ and 
\beq{}
\varkappa^2=4\Omega^2+r\frac{d\Omega^2}{dr}\equiv \frac{1}{r^3}\frac{d\Omega^2r^4}{dr}
\eeq
is the epicyclic frequency. For the power-law rotation $\Omega^2\sim r^{-n}$ the 
epicyclic frequency is simply $\varkappa^2/\Omega^2=4-n$.
In deriving these equations we neglected terms
$\sim(k_r/r), (k_z/r)$ compared to terms $\sim k^2$, 
see also the discussion in \cite{1978RSPTA.289..459A}.

\subsubsection{Energy equation}

To specify density perturbations $\rho_1/\rho_0$, the energy equation should be used. 
In the general case by varying the equation of state 
\Eq{e:eos} we obtain for entropy perturbations:
\beq{e:s1}
\frac{p_1}{p_0}=\frac{s_1}{c_V}+\gamma\frac{\rho_1}{\rho_0}\,.
\eeq
On the other hand, from the equation of state for ideal gas in the form 
$p=\rho {\cal R} T/\mu$, we find for small temperature perturbations  
we have:
\beq{e:T1}
\frac{p_1}{p_0}=\frac{\rho_1}{\rho_0}+\frac{T_1}{T_0}\,.
\eeq

Substitution of \Eq{e:s1} and \Eq{e:T1} into equations of motion \Eq{iur} and \Eq{iuz} immediately shows that the terms with $(p_1/p_0)k_r$ and $(p_1/p_0)k_z$ in the dynamical equations
are larger by factors $rk_r$ and $rk_z$ than terms with $p_1/p_0$ arisen from 
the energy equation. This means that in the energy equation we can set Eulerian
pressure perturbations equal to zero, as is usually assumed in the Boussinesq approximation, i.e. 
\beq{e:s11}
\frac{s_1}{c_V}+\gamma\frac{\rho_1}{\rho_0}=0\,.
\eeq
\beq{e:T11}
\frac{\rho_1}{\rho_0}=-\frac{T_1}{T_0}\,.
\eeq
We repeat again that the Eulerian pressure variations should be retained in the equations of motion  \eqn{iur}-\eqn{iuz}. 
\Eq{e:T11} implies that in the axially symmetric waves considered here the density variations are in counter-phase with temperature variations.

The viscous dissipative function $Q_\mathrm{visc}$ [erg~cm$^{-3}$~s$^{-1}$] 
can be written as $Q_\mathrm{visc}=\rho\nu\Phi$, where the function $\Phi$ in polar coordinates is
\begin{eqnarray}
\Phi=&2\left[
\myfrac{\partial u_r}{\partial r}^2+
\left(
\frac{1}{r}\myfrac{\partial u_\phi}{\partial \phi} +\frac{u_r}{r}
\right)^2+ 
\myfrac{\partial u_z}{\partial z}^2
 \right] \nonumber\\
&+\left[
r\frac{\partial}{\partial r}\myfrac{u_\phi}{r}+
\frac{1}{r}\frac{\partial u_r}{\partial \phi}\right]^2 +
 \left[\frac{1}{r}\frac{\partial u_z}{\partial \phi}\right]^2
 \nonumber\\
&+\left[\frac{\partial u_r}{\partial z} +
\frac{\partial u_z}{\partial r}\right]^2 -
\frac{2}{3}(\nabla\cdot\bm{u})^2\,.
 \nonumber\\
\end{eqnarray} 
All terms but one in this function are quadratic
in small velocity perturbations; this term has the form:
\beq{qterm}
\nu\rho \left( \frac{\partial u_\phi}{\partial r}-\frac{u_\phi}{r}\right)^2\,.
\eeq 
Writing for the azimuthal velocity $u_\phi=u_{\phi,0}+u_{\phi,1}$ (here for the purposes of this paragraph 
and only here we specially mark the unperturbed velocity with index 0, 
not to be confused with our notations $u_\phi$ for perturbed velocity in \Eq{iur}-\Eq{iuphi} above and below), we obtain for the viscous dissipation function 
\beq{Qv}  
Q_\mathrm{visc}=\nu\rho r \frac{d \Omega}{d r}\left[r\frac{d\Omega}{d
r}-2ik_r u_{\phi,1}-2\frac{u_{\phi,1}}{r}\right] + \hbox{quadratic\, terms}\,.
\eeq 
Here $\Omega=u_{\phi,0}/r$ is the angular (Keplerian) velocity of the unperturbed flow. 
The first term in parentheses describes the viscous energy release in the unperturbed Keplerian flow. For this unperturbed flow we have
\beq{}
\frac{\partial s_0}{\partial t}=\nu\mu \frac{[r(d \Omega/d r)]^2}{{\cal R}T_0}=\frac{9}{4}\nu\mu\frac{\Omega^2}{{\cal R}T_0}\,.
\eeq
Thus, the entropy of the unperturbed flow changes along the radius. However, on the scale of the order of or smaller than the disc thickness $z_0$, the entropy gradient can be neglected.  
The second term in \Eq{Qv} vanishes if $k_r=0$ (and then there are no 
viscous dissipation effects to linear order), therefore 
we will consider only two-dimensional transverse perturbations with $k_z\ne 0$, $k_r\ne 0$. 

We emphasize that in our analysis we neglect the background entropy gradients, which
can be present in real flows, i.e. we will consider the flow in the local neutral equilibrium. 
 This is done to exclude the effects of these gradients on
the evolution of small perturbations. As is well known, with inclusion of the background
entropy gradients, the Brunt-V\"ais\"al\"a frequencies arise. If their squares are positive, they stabilize perturbations.
If their squares are negative, they signal emergence of convection (see, for example, 
\cite{1998bhad.conf.....K}, for more detail).

The right side of the 
heat conductivity equation \Eq{e:kappat} for small temperature perturbations, with account for the dependence of the thermal conductivity coefficient on temperature and density $\kappa\sim T^c\rho^d$, in the linear order can be recast to the form
\begin{eqnarray}
&\kappa \Delta T+\nabla \kappa\cdot \nabla T = 
-\kappa k^2T_0\frac{T_1}{T_0}-\nonumber \\
&(c-d)\kappa T_0\frac{T_1}{T_0}ik_r\frac{1}{T_0}\frac{\partial T_0}{\partial r}-
(c-d)\kappa T_0\frac{T_1}{T_0}ik_z\frac{1}{T_0}\frac{\partial T_0}{\partial z}\,,
\end{eqnarray}
(here we have used the relation \eqn{e:T11}). It is useful to rewrite the right side of this equation in the form
\beq{}
-\kappa k^2T_0\frac{T_1}{T_0}\left(1+(c-d)i\frac{k_r}{k^2}\frac{1}{T_0}\frac{\partial T_0}{\partial r}+(c-d)i\frac{k_z}{k^2}\frac{1}{T_0}\frac{\partial T_0}{\partial z}\right)
\eeq	
to see that in so far as $\frac{1}{T_0}\frac{\partial T_0}{\partial r}\sim -1/r$, $\frac{1}{T_0}\frac{\partial T_0}{\partial z}\sim -1/z_0$ and that terms $\sim k_r/r$ should be neglected compared to terms $\sim k^2$, only the first term $\sim k^2$ and third term $\sim k_z/z_0$ should be retained in this equation.

Therefore, the energy equation \Eq{e:en} turns into
\beq{e:en1}
i\omega \frac{\rho_0{\cal R} T_0}{\mu}s_1=-2ik_r \nu\rho_0 r \frac{d \Omega}{d r}u_{\phi}
-\kappa k^2T_0\frac{T_1}{T_0}[E]\,,
\eeq
where we have introduced the correction factor
$$
[E]=1+i(c-d)\myfrac{k_z}{k^2}\frac{1}{T_0}\frac{\partial T_0}{\partial z}\,.
$$ 
Like in the linearized continuity equation $\nabla\cdot \bm{u}=0$, here we have neglected the term $u_{\phi}/r$.
The first term in the right side of \Eq{e:en1} corresponds to the energy generation in axially symmetric sheared flow due to viscosity, 
and the second term means the entropy 
perturbation smoothing due to heat conductivity. The first term $\sim (k_r/r)(u_{\phi,0}/u_s)^2\eta$, and the second term $\sim k^2\eta/\mathrm{Pr}$, where 
the Prandtl number is the ratio of the heat conductivity to dynamical viscosity 
coefficient. While the $k_r/r$ is small relative to $k^2$, the coefficient $(u_{\phi,0}/u_s)^2$ is very large for thin discs, and therefore the heat generation term should be
retained in the energy equation.

\subsection{Dispersion equation}

By substituting \Eq{e:s11} 
and \Eq{e:T11} into \Eq{e:en1}, we find the relation between the density variations and
$u_\phi$ in the Boussinesq limit 
with zero background entropy gradients: 
\beq{e:rho1uphi}
\frac{\rho_1}{\rho_0}\left(i\omega  + \frac{\kappa k^2[E]}{c_p\rho_0{\cal R}/\mu}\right)=
\frac{2ik_r \nu r (d \Omega/d r)}{c_p{\cal R}T_0/\mu}u_{\phi}\,.
\eeq 
Here $c_p=\gamma c_V=\gamma/(\gamma-1)$ is the specific heat capacity (per particle) at constant pressure. 

It is convenient to introduce the dimensionless Prandtl number:
\beq{e:Pre}
\hbox{Pr}\equiv \frac{\nu\rho_0 C_p}{\kappa}=\frac{\nu\rho_0({\cal R}/\mu) c_p}{\kappa}=
\frac{\nu \rho_0 ({\cal R}/\mu)}{\kappa}\frac{\gamma}{\gamma-1}\,.
\eeq
The Prandtl number defined by \Eq{e:Pre} for fully ionized hydrogen 
gas ($\gamma=5/3$), where the heat conduction is determined by light electrons, is quite low (see \cite{1962pfig.book.....S}):
\beq{e:Pr_spitzer}
\hbox{Pr}_\mathrm{e}\approx \frac{0.406}{20\cdot 0.4\cdot 0.225\cdot (2/\piup)^{3/2}}\myfrac{m_e}{m_p}^{1/2}\myfrac{5}{2}\approx 0.052\,.
\eeq
Note also that in this case $d=0$ and $c=5/2$ in the heat conductivity coefficient.

By expressing the heat conductivity coefficient $\kappa$ through kinematic viscosity coefficient $\nu$ using
\Eq{e:Pre} and after substituting \Eq{e:rho1uphi} into \Eq{iur}, we arrive at:
\begin{eqnarray}
&ik_r\displaystyle\frac{p_1}{\rho_0}
= (i\omega+\nu k^2[R])u_r +\displaystyle\frac{\varkappa^2}{(i\omega+\nu k^2[\Phi])}u_r\nonumber \\
&-\displaystyle\frac{1}{c_p}\displaystyle\frac{1}{p_0}\displaystyle\frac{\partial p_0}{\partial r}
\displaystyle\frac{\varkappa^2i\nu k_r (d\ln\Omega/d\ln r)}{(i\omega+\nu k^2[\Phi])(i\omega+\nu k^2[E]/\mathrm{Pr})}u_r\,.
\label{e:P1rho0r}
\end{eqnarray}

Now by substituting \Eq{e:rho1uphi} into \Eq{iuz} with account for \Eq{Bq}, we arrive at 
\begin{eqnarray}
&ik_z\displaystyle\frac{p_1}{\rho_0}
= -\displaystyle\frac{k_r}{k_z}(i\omega+\nu k^2[Z])u_r \nonumber \\
&-\displaystyle\frac{1}{c_p}\displaystyle\frac{1}{p_0}\displaystyle\frac{\partial p_0}{\partial z}
\displaystyle\frac{\varkappa^2i\nu k_r (d\ln\Omega/d\ln r)}{(i\omega+\nu k^2[\Phi])(i\omega+\nu k^2[E]/\mathrm{Pr})}u_r\,.
\label{e:P1rho0z}
\end{eqnarray}

Finally, by subtracting \Eq{e:P1rho0z} multiplied by $k_r$ from  \Eq{e:P1rho0r} multiplied by $k_z$  we arrive at the dispersion equation:
\begin{eqnarray}
\label{e:dispeq}
&(i\omega+\nu k^2[\Phi])
\left[(i\omega+\nu k^2[R])\displaystyle\frac{k_z^2}{k^2}+
(i\omega+\nu k^2[Z])\displaystyle\frac{k_r^2}{k^2}\right] \nonumber \\
&+\displaystyle\myfrac{k_z}{k}^2\varkappa^2\left[1-\displaystyle\frac{\gamma-1}{\gamma}
\displaystyle\frac{ik_r}{(i\omega+\nu k^2[E]/\hbox{Pr})}\left(A-\displaystyle\frac{k_r}{k_z}B\right)\right]=0\,,
\end{eqnarray}
where
\beq{e:A}
A\equiv \nu \frac{d\ln \Omega}{d\ln r}\frac{1}{p_0}\frac{\partial p_0}{\partial r}
\eeq
\beq{e:B}
B\equiv \nu \frac{d\ln \Omega}{d\ln r}\frac{1}{p_0}\frac{\partial p_0}{\partial z}
\eeq

The expression in the square brackets in \Eq{e:dispeq} 
above can be rewritten in the equivalent form:
\beq{}
\left[1+\frac{\gamma-1}{\gamma}
\frac{i\nu}{(i\omega+\nu k^2[E]/\hbox{Pr})}\frac{d\ln\Omega/d\ln r}{{\cal R}T_0/\mu}
\left(k_rg_{r,eff}-\frac{k_r^2}{k_z}g_z\right)\right]\,,
\eeq
where $g_{r,eff}=-1/\rho_0 (\partial p_0/\partial r)$ and 
$g_z=-1/\rho_0 (\partial p_0/\partial z)$ are the effective radial and 
vertical gravity accelerations in the unperturbed flow, respectively.
Clearly, the term $k_rg_{r,eff}\sim k_r/r$ is much smaller than 
$(k_r^2/k_z)g_z\sim (k_r^2/k_z)1/z_0$ and will be neglected in the further analysis. 
Note that if the dynamic viscosity coefficient is independent of temperature (i.e. $\alpha_{visc}=0$), correction factors $[R]=[\Phi]=[Z]=1$ and the first line in 
\Eq{e:dispeq} is simplified to $(i\omega +\nu k^2)^2$. We will see below that 
deviations of these correction factors from unity insignificantly affect 
the result of our analysis.

\section{Analysis of the dispersion equation}
\label{s:analysis}

\begin{figure*}
\begin{center}
\includegraphics[width=0.5\textwidth,height=10cm]{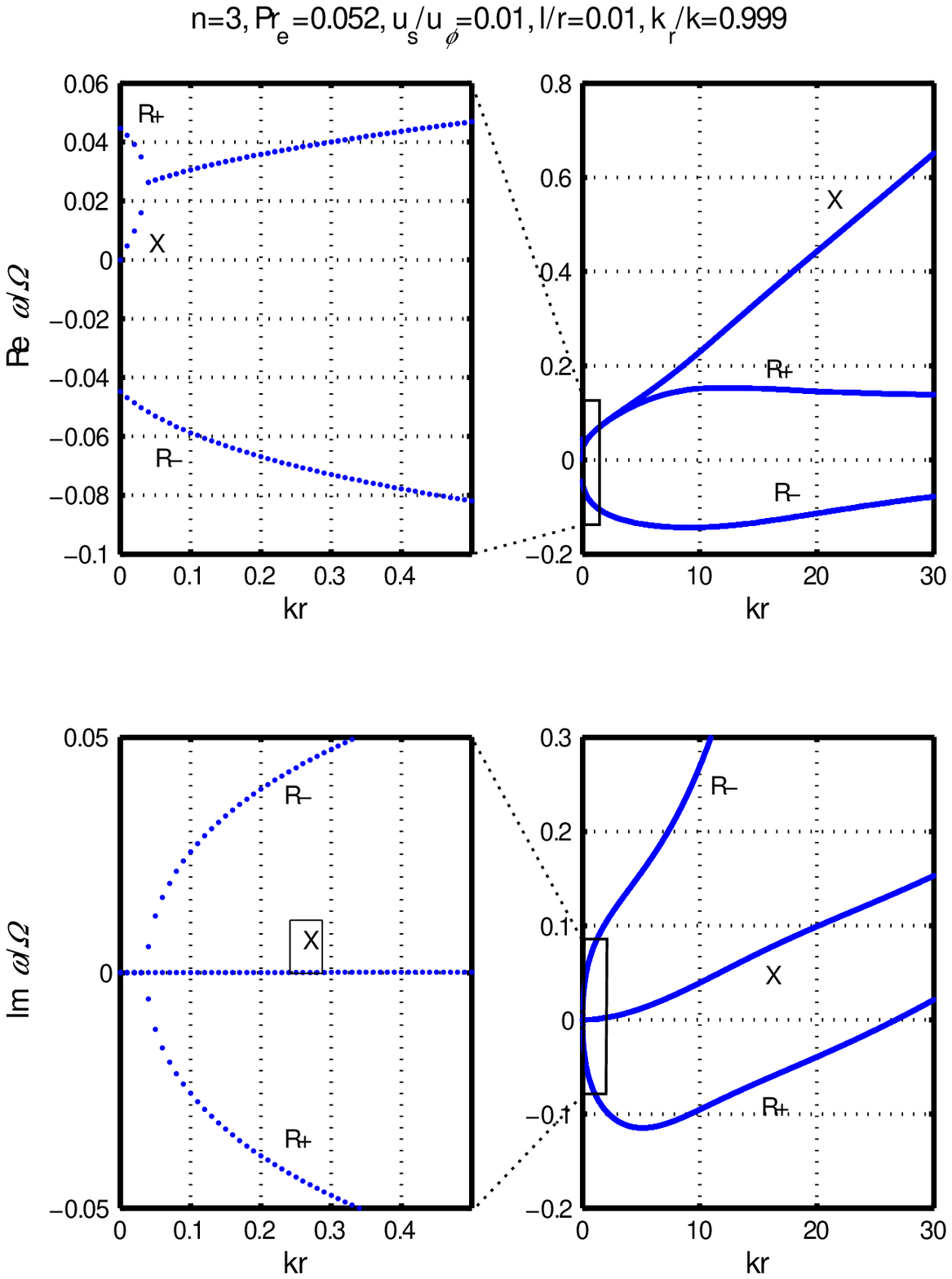}
\hfill
\includegraphics[width=0.225\textwidth,height=10cm]{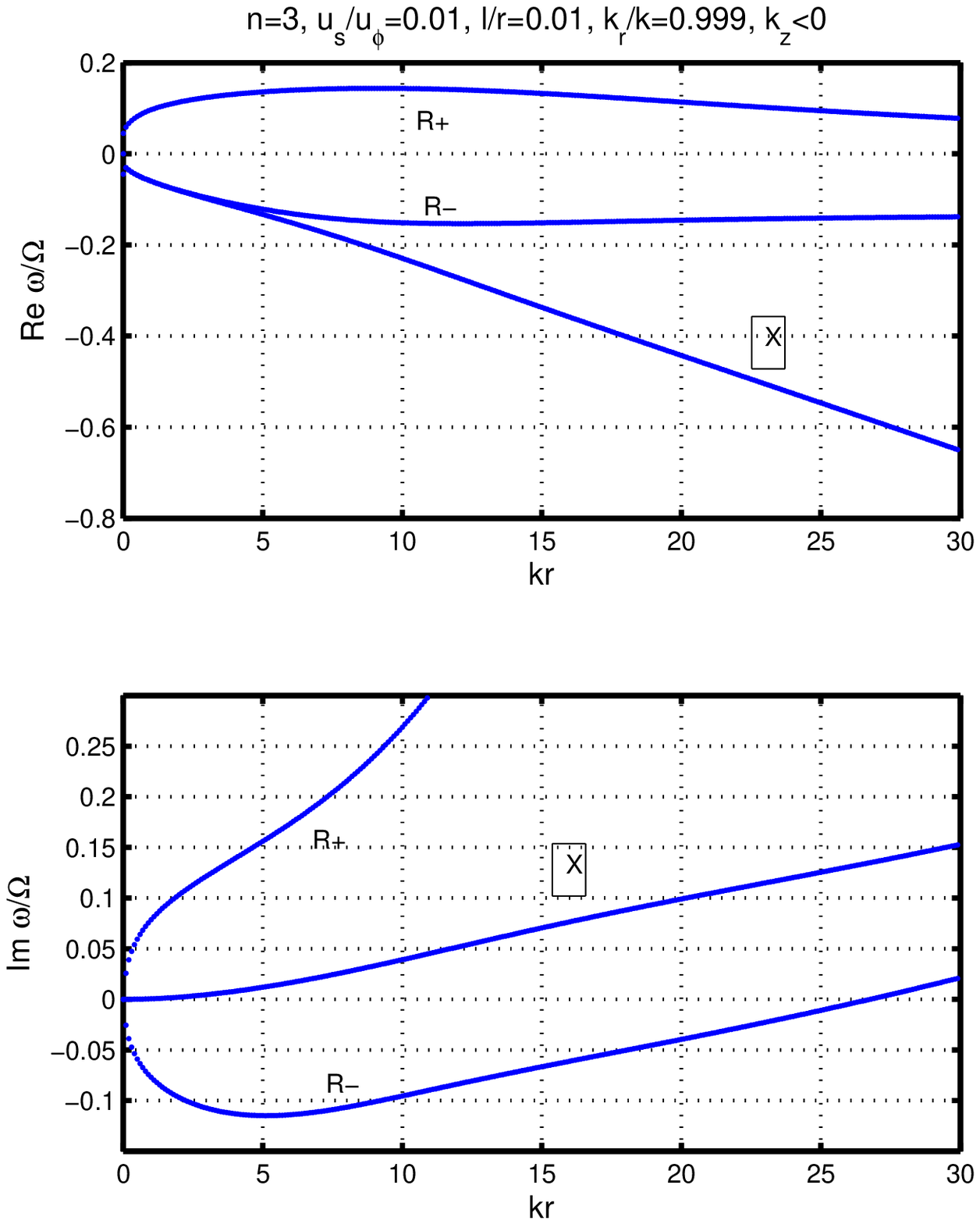}
\hfill
\includegraphics[width=0.225\textwidth,height=10cm]{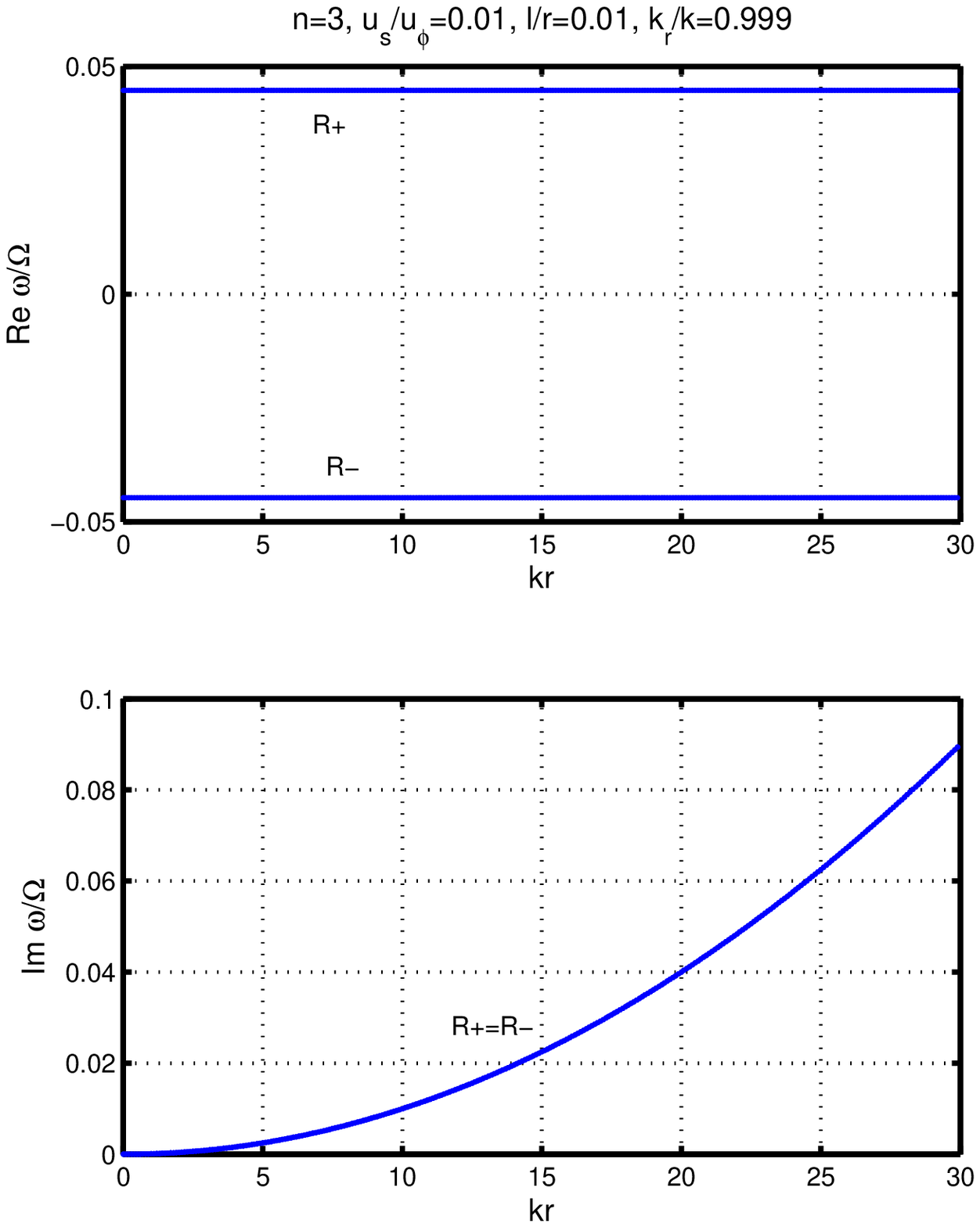}
\caption{Three modes of dispersion equation \protect{\eqn{e:dimlessde}} for electron thermal conductivity in fully ionized plasma (the Prandtl number Pr$_e$=0.052). R+ and R- mark two
Rayleigh modes, one of which (R+ if $k_z>0$ and R- if $k_z<0$) becomes viscously unstable. Upper row: Re$\tilde \omega$, bottom row: Im$\tilde\omega$. The first two
column show the case $k_r/k=0.999, k_z>0$. The left column zooms the region 
of small $kr$ to see the behaviour of three modes as $kr\to 0$: Re and Im parts of the X-mode always
starts from zero, while at $kr=0$ Re and Im parts of the Rayleigh modes is non-zero and zero, respectively.  The third column shows the case $k_r/k=0.999, k_z<0$: the imaginary part of the modes is unchanged and the real part changes sign. For comparison, the fourth column 
shows two (stable) Rayleigh modes as the solution of \Eq{e:Re}.}
\label{f:all}
\end{center}
\end{figure*}

First consider the limiting case where $A=B=0$ and $[R]=[\Phi]=[Z]=1$, 
i.e. the case where the 
viscous energy generation in the energy equation is ignored, but the viscosity is retained in the equations of motion. 
Then \Eq{e:dispeq} represents the well-known Rayleigh dispersion equation for viscous fluid:
\beq{e:Re}
(i\omega+\nu k^2)^2=-\myfrac{k_z}{k}^2\varkappa^2\,,
\eeq
which describes two Rayleigh modes. Depending on the sign of $\varkappa^2$ 
these modes are oscillating (if the epicyclic frequency $\varkappa^2>0$, the angular momentum increases with radius), 
or exponentially growing (the unstable Rayleigh mode) and exponentially decaying  
(if the epicyclic $\varkappa^2<0$, the angular momentum decreases with radius). The arising of the unstable 
Rayleigh mode corresponds to the classical Rayleigh criterion of instability of a shearing flow. 
As can be easily seen from \Eq{e:Re}, the viscosity stabilizes the unstable mode at short 
wavelengths (large $k$). Two solutions of \Eq{e:Re} for typical viscosity parameters discussed below and $\varkappa^2=\Omega^2$ (the Keplerian motion) are shown in the fourth column of Fig. \ref{f:all}. 

However, in the general viscous case where $A, B\ne 0$, the dispersion equation \Eq{e:dispeq} turns into 
a cubic equation, that is, the third mode arises (the X-mode) due to the 
viscous heating of the fluid. At non-zero $A$ and $B$, one of the Rayleigh 
modes (which is stable in the dissipationless case)
becomes exponentially unstable in a wide range of wavenumbers.
We stress that these modes remain stable for 
either $k_r=0$ or $k_z=0$. Indeed, the dispersion equation for 
perturbations with $k_r=0$ is reduced to \Eq{e:Re} above. For perturbations
with $k_z=0$, the dispersion equation turns into $i\omega+\nu k_r^2=0$, i.e.
is reduced to an exponentially decaying standing wave. 
We stress that in our case the larger viscosity, the higher 
instability increment.
This is opposite to
the situation where a poloidal magnetic field is present, when 
the magneto-rotational instability is developed: increasing viscosity 
decreases the MRI increment.
Clearly, there is no viscous instability of the Rayleigh modes
in the inviscid case ($\nu=0$) or in the shearless case (solid-body rotation with $n=0$).

\subsection{Dimensionless dispersion equation}
\label{ss:dimless}

For numerical analysis, the cubic dispersion equation \Eq{e:dispeq} can be  
conveniently rewritten in the dimensionless form. To do this, we multiply \Eq{e:dispeq}
through the factor $(i\omega +\nu k^2[E]/\mathrm{Pr})$, divide the obtained equation 
through $\Omega^3$ and introduce new dimensionless variables:
\beq{}
\tilde \omega \equiv \frac{\omega}{\Omega}\,,\quad kr\,, 
\quad \tilde \varkappa^2\equiv \frac{\varkappa^2}{\Omega^2}.
\eeq
The kinematic viscosity is $\nu=l u_s$, where $l$ is the effective mean free path of
ions, $u_s=\sqrt{\gamma{\cal R}T_0/\mu}$ is the characteristic velocity in the unperturbed flow which is about thermal velocity of ions, so the dimensionless combination $\nu k^2/\Omega$ becomes:
\beq{}
\frac{\nu k^2}{\Omega}=a(kr)^2\,.
\eeq
Here we have introduced the dimensionless coefficient 
\beq{e:a}
a\equiv\myfrac{u_s}{u_{\phi}} \myfrac{l}{r}\,.
\eeq
Formally, $1/a$ is the Reynolds number defined as 
Re$=(u_\phi r)/\nu$, but as we will see below, for a specified Reynolds 
number, different solutions are realized.

The vertical pressure gradient in 
coefficient $B$ in \Eq{e:dispeq} turns into 
\beq{}
\frac{1}{p_0}\frac{\partial p_0}{\partial z}\to -\frac{1}{z_0}\,,
\eeq
where $z_0$ is the characteristic disc height.
Using the relation for thin accretion discs 
\beq{e:z/r}
\frac{z_0}{r}=\sqrt{\Pi_1/\gamma}\myfrac{u_s}{u_\phi}
\eeq
(where the dimensionless coefficient $\Pi_1$ takes into account
the model vertical disc structure, see \cite{1998A&AT...15..193K}; in numerical calculation below we shall assume $\sqrt{\Pi_1/\gamma}=2$),
we obtain the dispersion equation in the dimensionless form:
\begin{eqnarray}
&(i\tilde\omega+a (kr)^2[\Phi])
\left[(i\tilde\omega+ a (kr)^2[R])\displaystyle\frac{k_z^2}{k^2}+
(i\tilde\omega+a (kr)^2[Z])\displaystyle\frac{k_r^2}{k^2}\right] \nonumber \\
%
&+\displaystyle\myfrac{k_z}{k}^2\tilde\varkappa^2
\left[1- i\displaystyle\frac{n}{2}(\gamma-1)\sqrt{\Pi_1/\gamma}\displaystyle\frac{(kr)\displaystyle\myfrac{l}{r}
\displaystyle\myfrac{k_r}{k}\displaystyle\myfrac{k_r}{k_z}}{i\tilde\omega+a(kr)^2[E]/\hbox{Pr}}\right]=0.
\label{e:dimlessde}
\end{eqnarray}
Here the dimensionless factors $[R]$, $[\Phi]$, $[Z]$ and $[E]$ have the form
\begin{eqnarray}
\label{e:[RPZE]}
&[R]=\left[1-i\alpha_{visc}\myfrac{k_z^2-k_r^2}{kk_z}\frac{1}{(kr)}\myfrac{u_\phi}{u_s}\right]\,, \nonumber\\
&[\Phi]=\left[1-i\alpha_{visc} \myfrac{k_z}{k}\frac{1}{(kr)}\myfrac{u_\phi}{u_s}\right]\,, \nonumber\\
&[Z]=\left[1-i2\alpha_{visc}\myfrac{k_z}{k}\frac{1}{(kr)}\myfrac{u_\phi}{u_s}\right]\,,\nonumber\\
&[E]=\left[1-i(c-d)\myfrac{k_z}{k}\frac{1}{(kr)}\myfrac{u_\phi}{u_s}\right]\,.
\end{eqnarray}

The inspection of \Eq{e:dimlessde} reveals the following properties of the solution:
\begin{itemize}
\item 
the solution should be independent on the radial direction of the perturbation wave since the radial component of the wave vector appears as $k_r^2$. Change of the sign of $k_z$ reverses the sign of the real part of the solutions (see the second and third column in Fig. \ref{f:all});
\item
in the limit of small viscosity, the second bracket in \Eq{e:dimlessde} 
becomes real in the first order, suggesting the stability. The decrease in
the viscous instability increment with decreasing $l/r$ is clearly 
seen in Fig. \ref{f:v001};

\item
for $u_s/u_\phi\sim 0.01$ (thin discs) and small $k_z/k\ll 1$, $k_r\sim 1$ and $kr\sim 10$, where the viscous instability appears (see below), the most appreciable correction is for the $[R]$-factor. However, in the dispersion equation \eqn{e:dimlessde} the term $(i\tilde\omega+a(kr)^2[R])$ is multiplied by the small value $(k_z/k)^2$, and therefore the effects from the correction factors $[R]-[E]$ on the solution of the dispersion equation should be not significant, as indeed we found to be the case.

\end{itemize}

This dimensionless dispersion equation for $\tilde\omega$ as a function of 
the dimensionless wavenumber $(kr)$ is to be solved for different values of the dimensionless parameters: the Prandtl number Pr, which characterizes the effect of thermal conductivity, 
$l/r$ and $u_s/u_\phi$, which describe the viscosity, and $k_r/k$, which 
determines the direction of the wave (evidently, $(k_z/k)^2=1-(k_r/k)^2$). 

\subsection{Numerical solution of the dispersion equation}
\label{ss:num}

In principle, it is possible to carry out analytical investigation of 
the properties of the solutions of the cubic equation \Eq{e:dimlessde}, e.g. in a way similar to study of MRI modes by \cite{2008ApJ...684..498P}. However, the main aim of the present paper is to show the existence of the viscous instability 
in shearing flows, therefore we will numerically solve \Eq{e:dimlessde} for 
different representative parameters.
Everywhere below in this Section we shall consider the phenomenologically important 
Keplerian case with $n=3$ and $\tilde\varkappa=1$. This does not restrict
our analysis, since the instability persists at any $n$ but $n=0$ (see the next Section). 

\subsubsection{Case of electron heat conductivity}

\begin{figure}
\begin{center}
\centerline{\includegraphics[width=0.5\textwidth]{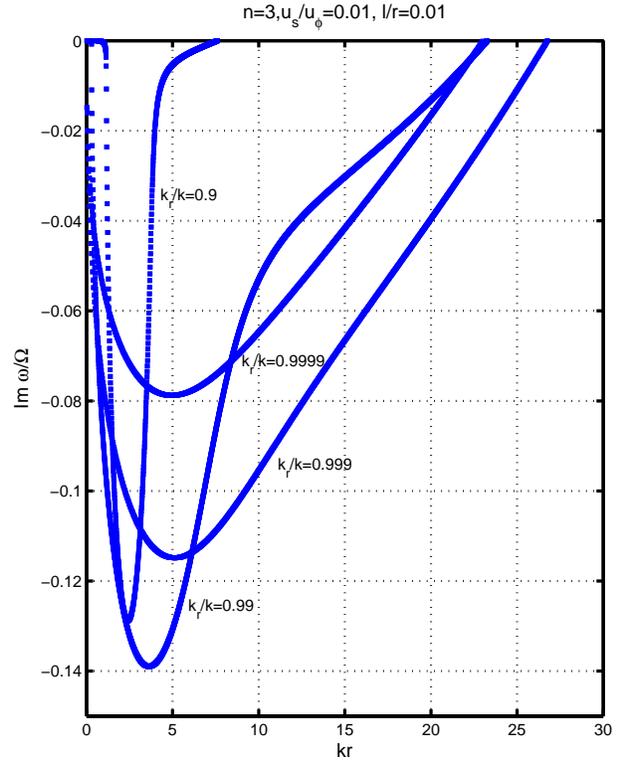}} 
\caption{Imaginary part of the viscously unstable mode R+ in fully ionized gas
with electron heat conductivity (Pr$_e$=0.052) and viscosity parameters 
$u_s/u_\phi=0.01, l/r=0.01$ 
for four values of the wave vector $k_r/k=0.9$, 
0.99, 0.999, and 0.9999.}
\label{f:kr}
\end{center}
\end{figure}

We start with the electron heat conductivity in a fully ionized plasma.
We remind that in this case the Prandtl number is Pr$_e$=0.052, the dynamical viscosity coefficient is
$\eta\sim T^{5/2}$, the heat conductivity coefficient is $\kappa\sim T^{5/2}$, so that $\alpha_{visc}=5/2$, $c=5/2$ and $d=0$ in \Eq{e:[RPZE]}.
Fig. \ref{f:all} 
shows the real (upper panels) and imaginary (bottom panels) 
parts of three solutions of the cubic dispersion equation
\Eq{e:dimlessde} $\tilde\omega$ as a function of the dimensionless wavenumber $kr$. All three solutions of this equation are complex since the dispersion equation has complex coefficients. 
The dimensionless parameters are 
$u_s/u_\phi=0.01$ (thin discs),  
$l/r=0.01$ (the maximum possible free-path length of ions, 
not to exceed the disc thickness), $k_r/k=0.999$ (the direction of perturbations with
increment close to maximal one for these parameters, see Fig. \ref{f:kr}). 
Two Rayleigh modes
modified by viscosity are marked as R+ and R-, according to the sign of their real parts at $kr\to 0$. The first two columns show the solutions for $k_r=0.999$ and 
positive $k_z>0$ and negative $k_z<0$, respectively. It is seen that the sign of $k_z$ determines which of the Rayleigh modes, R+ or R-, becomes unstable. It is also seen the unstable mode is that which has the real part intersecting with the new X-mode (the latter is always stable, i.e. has a non-negative imaginary part, representing an oscillating wave).
The unstable Rayleigh mode has a non-zero increment  
already for long perturbations with $kr \to 0$. It has a maximum increment of $\sim 0.1$ at $kr\approx 5$ and is stabilized by viscosity for $kr\gtrsim 25$.

Fig. \ref{f:kr} illustrates the effect of changing the perturbation propagation 
wavevector value
$k_r/k$ in the range from 0.9 to 0.9999. It is seen that at $k_r/k=0.999$ the instability increment is about maximum (we did not investigate the exact value of $k_r/k$ for maximum increment, which, if necessary, can be straightforwardly done by differentiating the dispersion equation with respect to $k_r/k$ and equating the result to zero).

\begin{figure*}
\begin{center}
\includegraphics[width=0.3\textwidth,height=7cm]{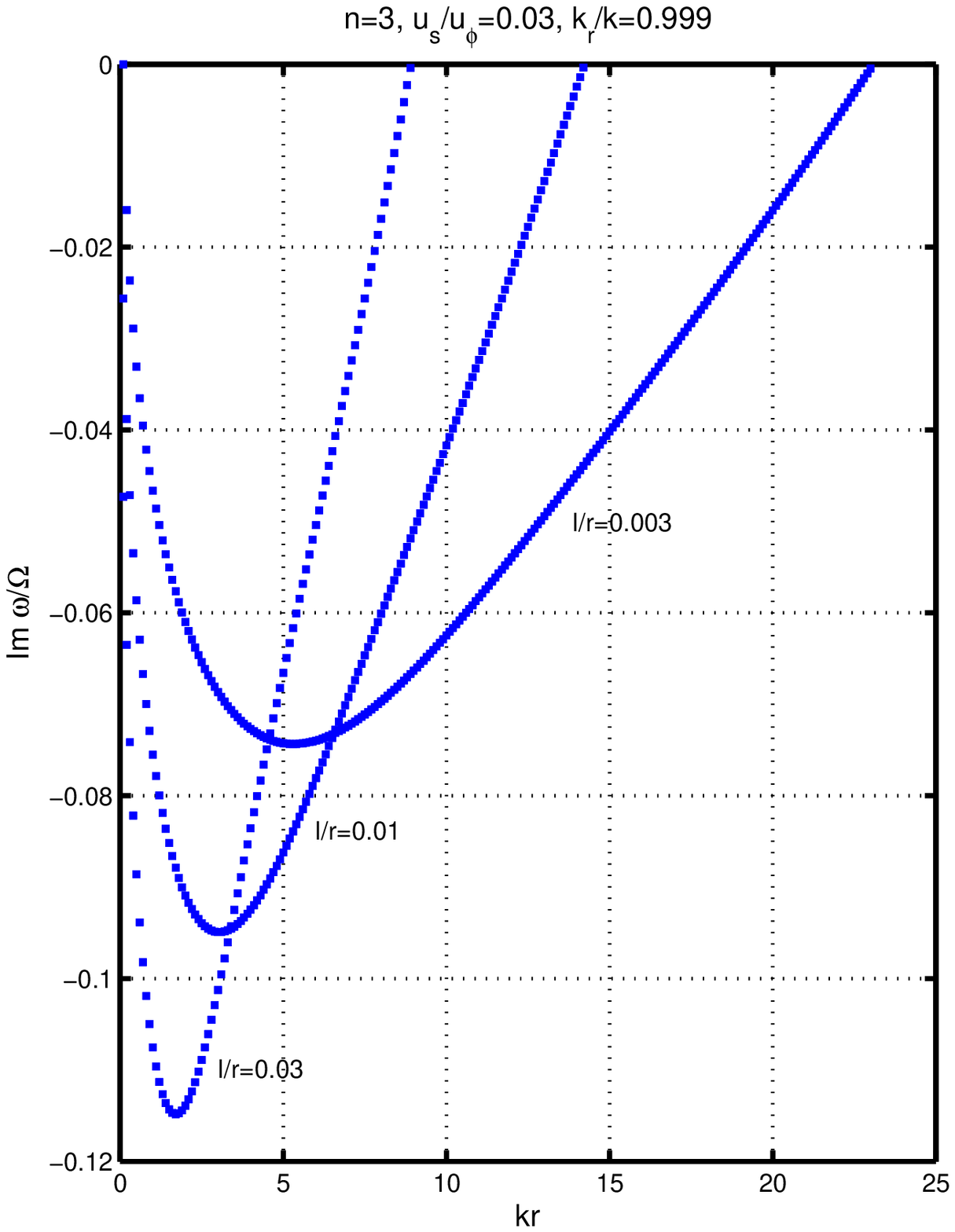}
\hfill
\includegraphics[width=0.3\textwidth,height=7cm]{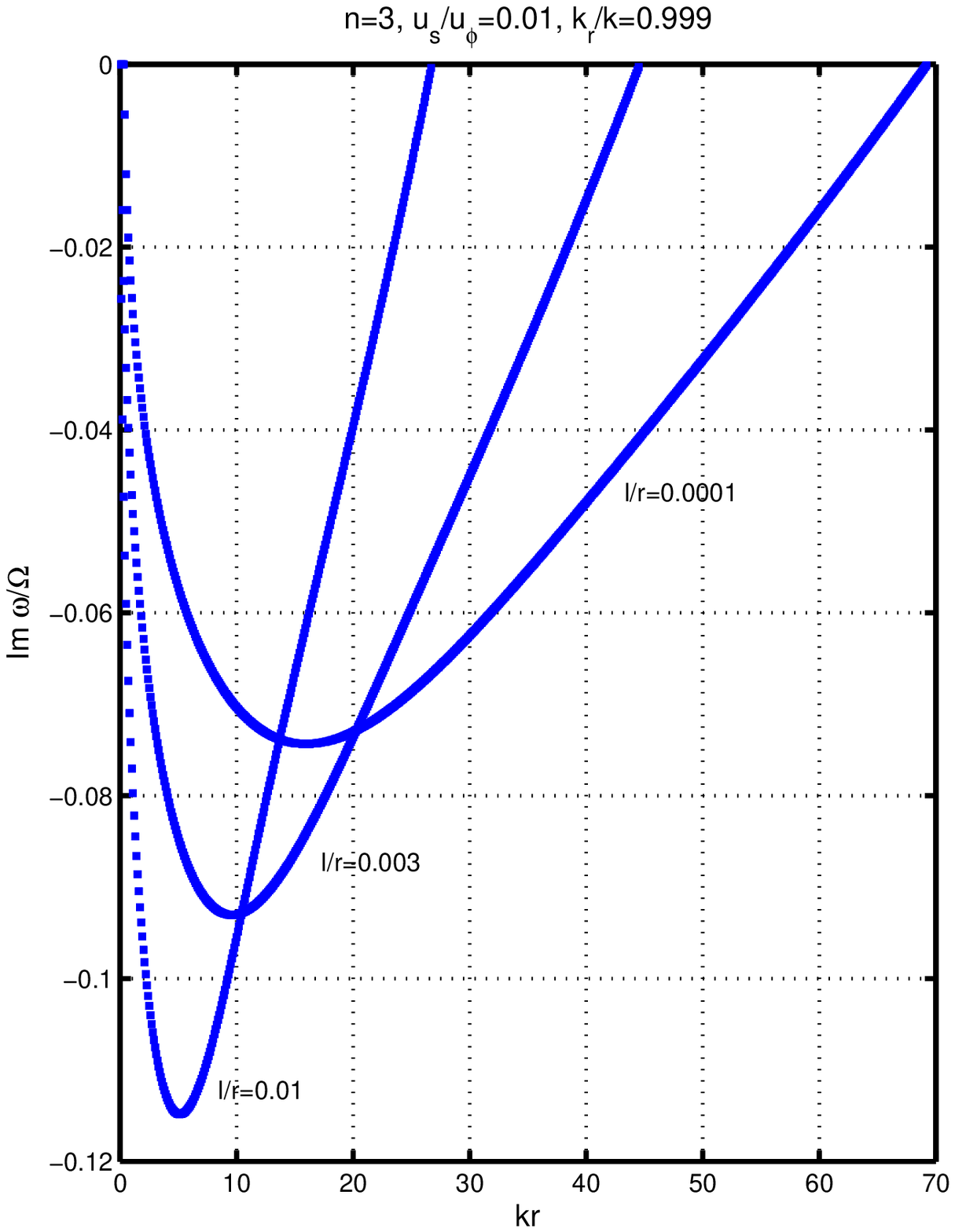}
\hfill
\includegraphics[width=0.3\textwidth,height=7cm]{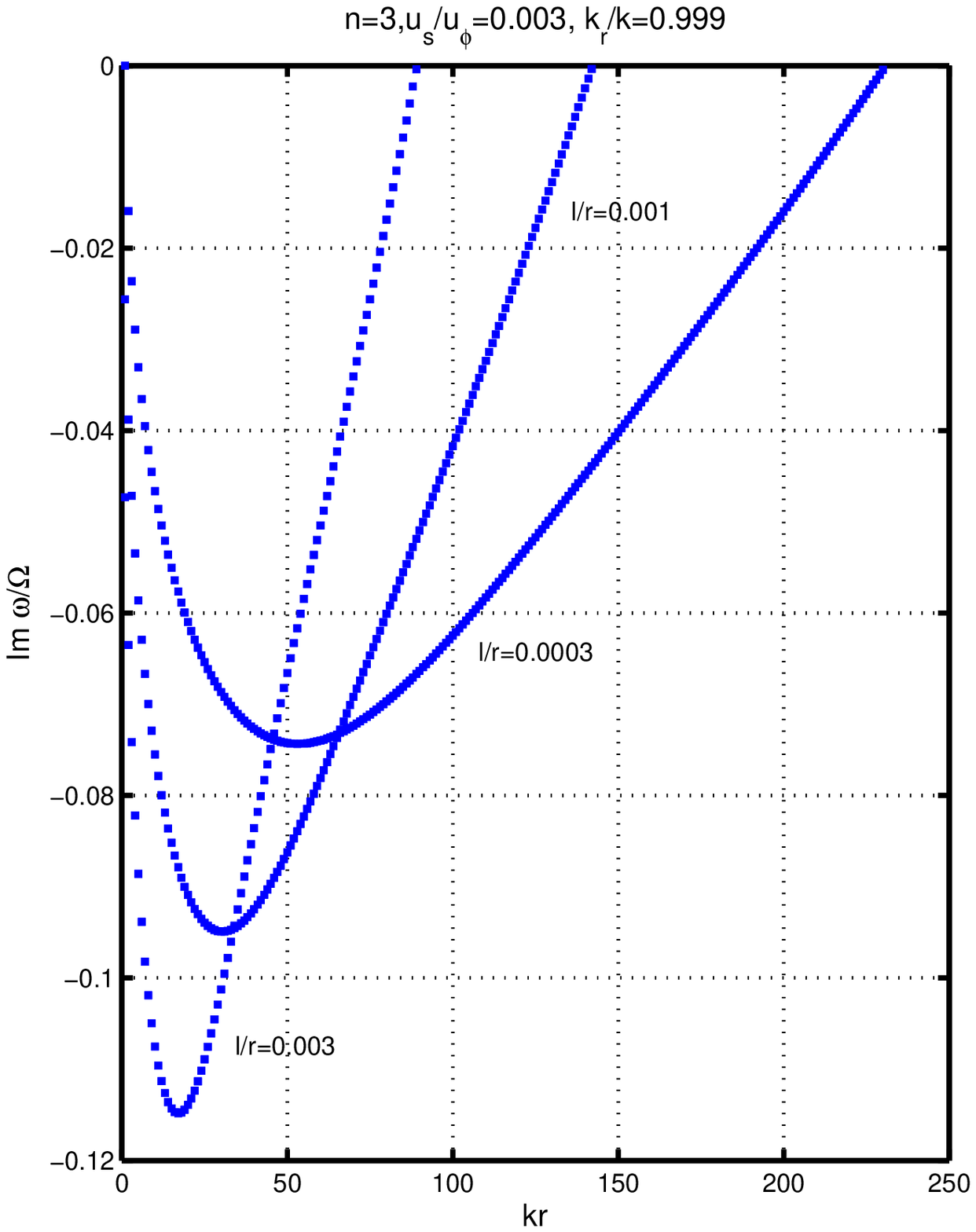}
\caption{Almost precise self-similarity of the solution for different disc thickness parameters $u_s/u_\phi$. Shown are three cases  for Pr$_e$=0.052, $k_r/k=0.999$ and (from left to right) $u_s/u_phi=0.03, 0.01, 0.003$ with different values of $l/r$. It is seen that the combination $a(kr)^2$ is almost exactly the same for all three figures.}
\label{f:v001}
\end{center}
\end{figure*}

Fig. \ref{f:v001} shows the unstable Rayleigh R+ mode behaviour with changing the viscosity 
parameter $l/r$ and other parameters fixed as in Fig. \ref{f:kr}. It is seen that 
diminishing the particle free-path length from the maximum possible value ($l/r=0.01$ in this case) 
by an order of magnitude decreases the Rayleigh R+ mode instability increment by 
about two times, but 
increases the instability interval from $kr\simeq 25$ to $kr\simeq 70$.

Fig. \ref{f:v001} also illustrates the effect of increasing or decreasing the disc thickness 
by three times ($u_s/u_\phi=0.03$ and $u_s/u_\phi=0.003$, respectively). 
The imaginary part of the unstable R+ mode is shown for 
three values of the viscosity parameter $l/r=0.03$, 
0.01 and 0.003. Note the close similarity (almost identity) of the curves to those in 
Fig. \ref{f:v001}, but the stretching of the $kr$ variable by three times. This reflects
an almost self-similarity of the dispersion equation \Eq{e:dimlessde} with respect to 
the dimensionless viscosity $a(kr)^2=(u_s/u_\phi)(l/r)(kr)^2$.

\subsubsection{Case of radiative heat conductivity}
\label{ss:radcond}

For a mixture of electrons and photons, the heat conductivity can be
characterized of an effective Prandtl number defined as 
\beq{e:Preff}
\frac{1}{\hbox{Pr}}\equiv 
\frac{1}{\hbox{Pr}_\gamma}+\frac{1}{\hbox{Pr}}
=\frac{1}{\hbox{Pr}}\left(1+\frac{\hbox{Pr}}{\hbox{Pr}_\gamma}\right)
=\frac{1}{\hbox{Pr}_{e}}\left(1+\frac{q_\gamma}{q_e}\right),
\eeq
where the heat flux due to electrons is
\beq{e:qe}
q_e=-\frac{1}{3}u_sl_en_e \nabla(k_\mathrm{B}T)
\eeq
and the heat flux due to photons is 
\beq{e:qg}
q_\mathrm{\gamma}=-\frac{1}{3}cl_\gamma  \nabla(a_\mathrm{r}T^4)
\eeq
where $a_\mathrm{r}$ is the radiation constant.
Therefore, 
\beq{e:qgqe}
\frac{q_\gamma}{q_e}\simeq \beta \frac{c}{u_s}\frac{z_0}{\tau}\sigma_{ei} n_e
\eeq 
where $\beta\equiv p_\gamma/p_\mathrm{gas}$ is the radiation to gas pressure ratio, 
$\tau$ is the effective optical thickness of the disc and $\sigma_{ei}$ is the electron-ion
interaction cross-section. Noticing that $n_e\sigma_{ei} u_s=\nu_{ei}=\Omega_{ei}/2\piup$ is the electron-ion
collisional frequency, \Eq{e:qgqe} can be recast into the form
\beq{e:qgqe1}
\frac{q_\gamma}{q_e}\simeq \frac{\beta}{2\piup}\frac{\Omega_{ei}/\Omega}{\tau} \frac{c}{u_s}\,.
\eeq

\begin{figure}
\begin{center}
\centerline{\includegraphics[width=0.5\textwidth]{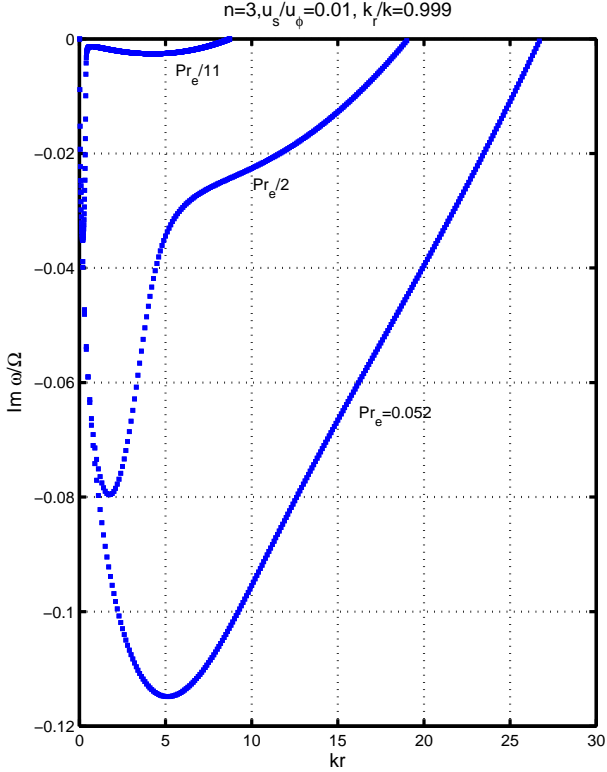}} 
\caption{Imaginary part of the unstable mode R+ for $u_s/u_\phi=0.01$, 
$l/r=0.01$, $k_r/k=0.999$ and different values of the 
effective Prandtl number \eqn{e:Preff}, illustrating the effect 
of radiative conductivity growth.}
\label{f:rad}
\end{center}
\end{figure}

Fig. \ref{f:rad} shows the effect of decreasing the effective Prandtl number
due to increase of the radiation heat conductivity. Two cases with Pr=Pr$_e$/2, 
Pr$_e$/11 are shown in comparison with the case of electron heat conductivity only. 
It is seen that the radiation conductivity in fully ionized plasma strongly decreases
(but does not vanish) the instability increment.

\subsubsection{Case of cold neutral gas}
\label{ss:cold}

Let us discuss the case of cold neutral gas. In this case the Prandtl number  Pr$_n$=2/3 according to simplified kinetic theory 
\citep{Hirschfelder_al54} and the heat conductivity 
coefficient $\kappa\sim T^{1/2}$ ($c=1/2$, $d=0$) \citep{1962pfig.book.....S}.

\begin{figure}
\begin{center}
\centerline{\includegraphics[width=0.5\textwidth]{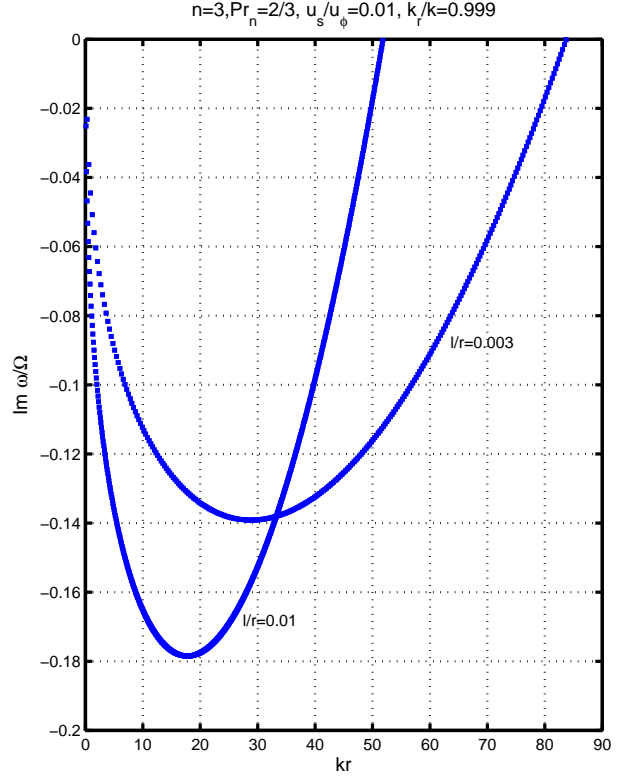}} 
\caption{Imaginary part of the unstable mode R+  in the case of ideal neutral hydrogen gas with Pr$_n=2/3$ for $k_r/k=0.999$, $u_s/u_\phi=0.01$,  and viscosity parameter $l/r=0.01$.}
\label{f:cold}
\end{center}
\end{figure}

Fig. \ref{f:cold} shows the imaginary part of the unstable mode R+ for 
the standard parameters $u_s/u_\phi=0.01$, $k_r/k=0.999$ used above in the case of ideal neutral hydrogen gas with 
Pr$_n=2/3$ and the viscosity parameters $l/r=0.01$ and $l/r=0.03$. As above, the decrease in the particle free-path length widens the instability wavelength interval and decreases the instability increment. 
In this case, the instability increment is maximum 
at $kr\simeq 17$ and is about 0.18, almost two times as large as in the case of 
the purely electron heat 
conductivity in fully ionized gas discussed above. Therefore, the viscous instability 
turns out to be the most strong in the case of cold neutral gases.

\section{Sheared flows with non-Keplerian rotation}

Here we discuss the behaviour of the viscously unstable Rayleigh mode 
in flows with possible non-Keplerian rotation 
(i.e where $\Omega^2\propto r^{-n}$ and $n\ne 3$). 
The solid-body rotation case with $n=0$ was already discussed above. In that case there is no shear and the coefficients $A=B=0$ in \Eq{e:dispeq} 
but the viscosity remains in equations of motion (see the discussion at 
the beginning of Section \ref{s:analysis}). 

\begin{figure}
\begin{center}
\centerline{\includegraphics[width=0.5\textwidth]{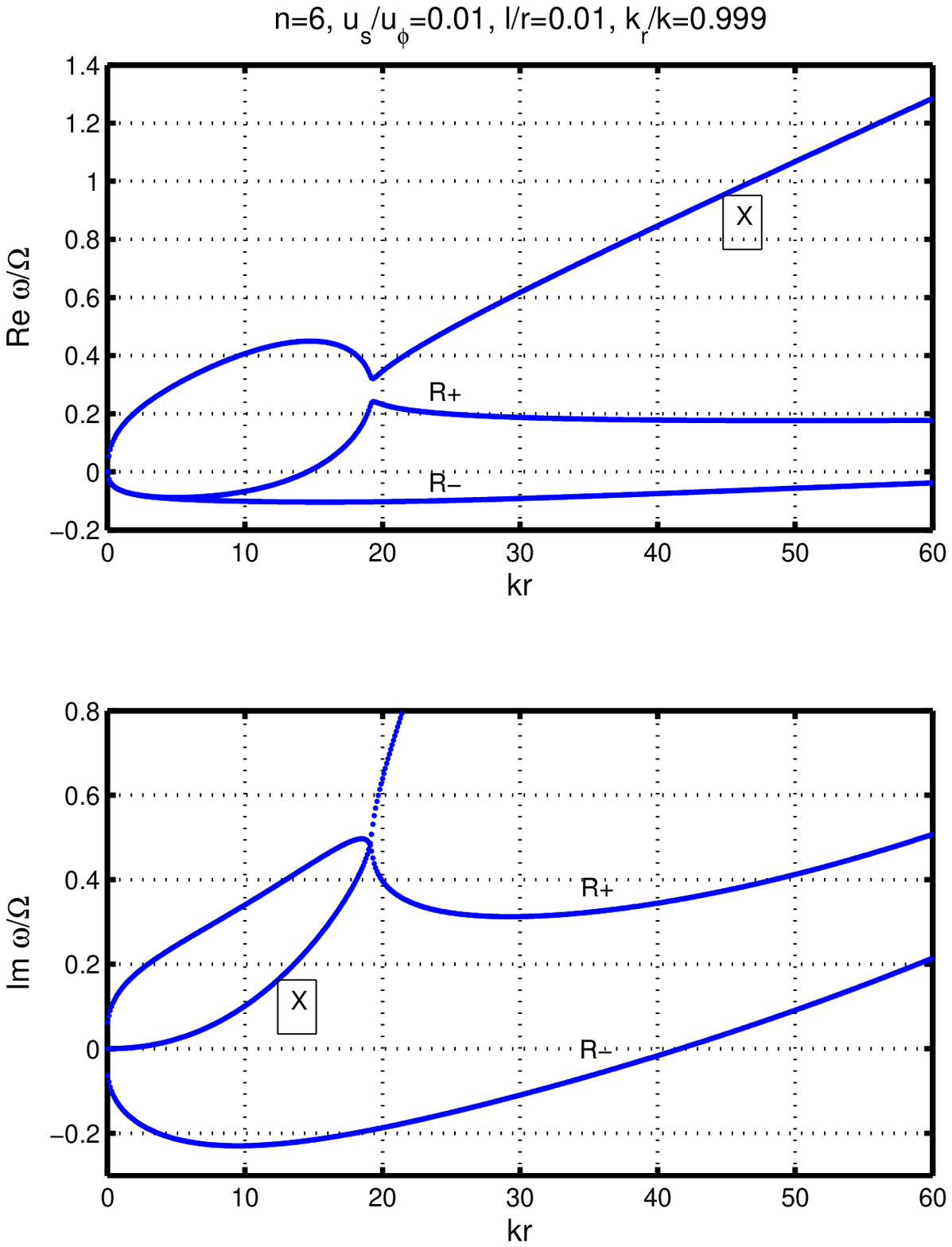}} 
\caption{Three modes of dispersion equation \Eq{e:dimlessde} in the case of
a Rayleigh-unstable flow with $n=6$.}
\label{f:n6}
\end{center}
\end{figure}

The case of a Rayleigh-unstable flow with $n>4$ (i.e. with specific angular momentum  decreasing outward) is shown in Fig. \ref{f:n6}.  
Here the R- mode is unstable (unlike in the Keplerian case with $k_z>0$) in a wide range of $kr$ with non-zero negative imaginary part at $kr\to 0$
and the maximum instability increment $\sim 0.2$.

\begin{figure}
\begin{center}
\centerline{\includegraphics[width=0.5\textwidth]{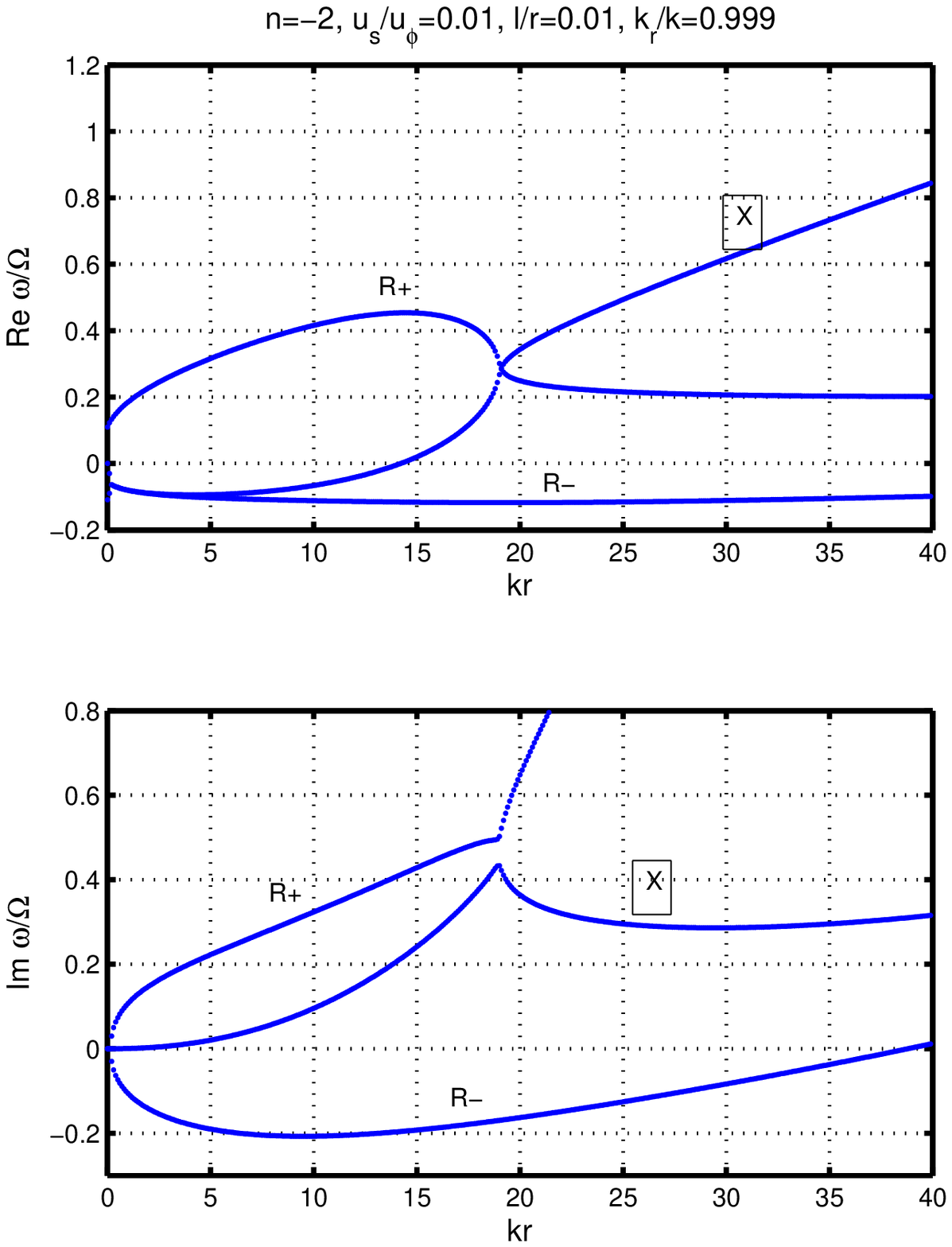}} 
\caption{Three modes of dispersion equation \Eq{e:dimlessde} in the case of
a flow with angular velocity linearly increasing with radius $\Omega\sim r$ ($n=-2$),
which is stable against MRI.}
\label{f:n-2}
\end{center}
\end{figure}

Now consider a flow with increasing angular velocity with radius, i.e. with $n<0$.
As is well known, such flows are MRI-stable \citep{Velikhov59, 1960PNAS...46..253C}. 
However, the viscous instability discussed in this paper persists in this case 
(see Fig. \ref{f:n-2}). Like in the case with $n=6$, the R- mode is unstable in a wide range of $kr$.

Finally, the flow with constant angular momentum ($n=4$) corresponds to $\varkappa=0$ 
and deserves special consideration. 
Such flows can be realized in various astrophysical situations, e.g. in quasi-spherical
accretion with angular momentum onto compact stars \citep{2012MNRAS.420..216S}.
If the correction factors $[R]-[E]$ \eqn{e:[RPZE]}
were ignored, pure decay of perturbations 
due to viscosity would take place, $\omega = i\nu k^2$. However, if they are taken into account, the solution of \Eq{e:dimlessde} is
\beq{}
\omega=i\nu k^2+\alpha_{visc}\nu k^2\frac{k_z}{k}\frac{1}{(kr)^2}\frac{u_\phi}{u_s}\,,
\eeq
representing decaying oscillations with frequency $\sim \alpha_{visc}\Omega_K (lk_z)$, which can be much smaller than the Keplerian one.

\section{Discussion}
\label{s:disc}

\subsection{Justification of the approximation of incompressibility}

As is well known (see \cite{1959flme.book.....L}), 
the approximation of incompressibility requires the characteristic time 
of the density change in a fluid to 
satisfy the relation $\tau\gg L/c_s$, where $L$ is the 
characteristic scale of the problem. For perturbations with 
the characteristic frequency $\omega=2\piup/\tau$ and 
wavenumber $k=2\piup/L$ this general relation yields $\omega\ll kc_s$, 
and in the thin discs with $c_s\sim \Omega z_0$ we obtain the condition 
of the incompressibility in the form 
\beq{}
kz_0\gg \frac{\omega}{\Omega}\,.
\eeq 
Writing $kz_0=(kr)(z_0/r)=(kr)(u_s/u_\phi)$, this condition becomes
\beq{}
(kr)\gg \frac{(\omega/\Omega)}{(u_s/u_\phi)}\,.
\eeq
For thin discs with $u_s/u_\phi\sim 0.01-0.03$ and for the found
mode frequencies $\omega \lesssim 0.1 \Omega$ we see that the assumption of the incompressibility is valid for modes with $(kr)\gg 3-10$. This implies that 
in the range $(kr)\sim 10-50$ where the viscous instability considered here 
reaches maximum increments (especially in the case of cold neutral gases) 
the assumption of incompressibility is justified and 
sound wave modes can be ignored.

\subsection{Damping by entropy gradients}

So far we have ignored the possible radial and vertical entropy gradients, i.e.
have dealt with locally adiabatic flow. 
As is well known (see, e.g., \cite{1998bhad.conf.....K}), the presence of 
non-zero 
entropy gradients $s_r\equiv\partial s/\partial r$ 
and $s_z\equiv\partial s/\partial z$  
can stabilize instabilities.  
For example, if the vertical temperature gradient in a flow is non-adiabatic, 
$dT/dz < dT/dz_{ad}=g_z/C_p$, the restoring gravity force would suppress
the development of convection, leading to an oscillatory vertical 
motion of a gas parcel 
with the Brunt-V\"ais\"al\"a frequency $N_z$. Qualitatively, it is expected that if this frequency is larger than the instability increment, the perturbation amplitude will not increase.    
To quantify this, 
we introduce the entropy gradients into the right-hand side of energy equation 
\eqn{e:en}, and arrive at the modified dispersion equation:
\begin{eqnarray}
\label{e:dispeqN}
&(i\omega+\nu k^2[\Phi])
\left[(i\omega+\nu k^2[R])\displaystyle\frac{k_z^2}{k^2}+
(i\omega+\nu k^2[Z])\displaystyle\frac{k_r^2}{k^2}\right] \nonumber \\
&+\displaystyle\myfrac{k_z}{k}^2\varkappa^2\left[1
+\displaystyle\frac{(i\omega+\nu k^2[\Phi])}{(i\omega+\nu k^2[E]/\hbox{Pr})}
\displaystyle\frac{\left(N_r-\displaystyle\frac{k_r}{k_z}N_z\right)^2}{\varkappa^2}\right. \nonumber \\
&-\left.\displaystyle\frac{\gamma-1}{\gamma}
\displaystyle\frac{ik_r}{(i\omega+\nu k^2[E]/\hbox{Pr})}\left(A-\displaystyle\frac{k_r}{k_z}B\right)\right]=0\,.
\end{eqnarray}
Here $N_r^2=-S_rg_r$ and $N_z^2=-S_zg_z$ are the Brunt-V\"ais\"al\"a frequencies. 
It is seen that it is the vertical Brunt-V\"ais\"al\"a frequency $N_z$ that mostly affects the results, the radial oscillations being suppressed by small 
factor $k_z/k_r$. We find that in the case of Keplerian rotation of ionized ideal gas with Pr$_e$=0.052 and
$k_r/k=0.999$ the viscous instability discussed above disappears for $N_z/\Omega\gtrsim 0.3$. Neutral gas with Pr$_n$=2/3 is stabilized if $N_z\gtrsim 0.35$. For example, for a polytropic thin accretion discs with vertical structure described by the polytropic index $n'$ , $P= K\rho^{1+1/n'}$, discussed
in \cite{1998A&AT...15..193K}, $N_z^2=2z^2(n'-3/2)\Omega_K^2/(1-z^2)$, where $\Omega_K$ is the Keplerian rotation frequency. The Brunt-V\"ais\"al\"a frequency $N_z$ averaged over the disc height $z_0$ is $\langle N_z/\Omega_K\rangle=\sqrt{(n'-3/2)/2}$. Therefore, the value $N_z=0.3$ corresponds to a polytropic index $n'\simeq 1.7$.
Of course, realistic flows can be not polytropic, and therefore effects of
the entropy gradients on the viscous instability should be investigated separately in each particular case.

\section{Summary and conclusion}
\label{s:concl}

In the present paper we have performed a linear local WKB analysis 
of time evolution of 
small axisymmetric perturbations in sheared hydrodynamic laminar flows. As a simplification, 
we have used the Boussinesq approximation for the description of the perturbations, but
included the viscous dissipation and heat conductivity terms in the energy equation. 
This procedure led us
to a third-order algebraic dispersion equation with complex coefficients (see \Eq{e:dimlessde}). 
The inclusion of these terms makes one of the Rayleigh modes (with positive or negative real part depending on the sign of the wavevector component $k_z$) unstable 
for long-wave perturbations for locally adiabatic case (i.e. ignoring local entropy gradients). The new X-mode of this cubic equation is found to be always stable (i.e. has a positive imaginary part). 

We have studied numerically the behaviour of the unstable Rayleigh mode in the most interesting case of 
thin Keplerian accretion discs for different values of 
the viscosity (which is parametrized by the mean free-path length of ions), disc thickness
(which is described by the ratio of the sound velocity to the unperturbed tangential velocity 
in the flow),  the directions of
the perturbation propagation (which is described by the ratio of wave vector components $k_r/k_z$), and the Prandtl numbers (which describe the heat conductivity 
effects). We have found that the value of 
heat conductivity mostly affect the instability increment, 
which is found to be maximum $\sim 0.2$ of the local Keplerian frequency in the case of cold
neutral gas with the highest value of the Prandtl number Pr$_n=2/3$ (see Fig. \ref{f:cold}). In the fully ionized gas characterized by the Prandtl number Pr$_e=0.052$
for purely electron heat conductivity, 
the instability increment is about $0.1$ and decreases with increasing the role of the 
radiation heat conductivity (Fig. \ref{f:rad}). The instability increment does not sensitive to the 
direction of propagation of perturbations (the sign of wavenumbers $k_r$ and $k_z$)
(Fig. \ref{f:all}) and
persists as long as shear and viscosity are present in the flow and the flow
is not iso-momentum when the epicyclic frequency vanishes, i.e. for any 
law of the angular momentum $\Omega^2\sim r^{-n}$ (see Fig. \ref{f:n6} and Fig. \ref{f:n-2}).

In the presence of viscous dissipation, the instability arises when the pressure gradients along radial or vertical coordinates are non-zero, suggesting its convective nature: the heat 
generation in a sheared viscous flow in the gravity field of the central star makes
the flow convectively unstable. Different aspects of 
convection in cold accretion discs, especially
suitable for the physics of protoplanetary discs, has been addressed 
in many papers, starting from the pioneer paper by \cite{1980MNRAS.191...37L} 
(see also \cite{1992ApJ...388..438R, 2010MNRAS.404L..64L}, and references therein).

We show that the incompressibility 
approximation is  applicable to describe small perturbations
in thin accretion discs with not very long wavelength ($kr\gg 3-10$).
At longer wavelengths, acoustic perturbations should be taken into account. 
On the other hand, the local WKB analysis is applicable only for $kr\gg 1$.
Thus, the found instability with maximum increment at $kr\sim 10-50$
seems to be robust under our assumptions.

Thus we conclude that the viscous instability of one of 
the classical Rayleigh mode discovered in the present paper
may be a seed for the development of turbulence in sheared flows which are hydrodynamically
stable according to the classical Rayleigh criterion (i.e. in which the angular momentum increases
with radius), or stable against MRI (e.g. flows with angular velocity increasing with radius). This instability is certainly worth investigating further.

\section{Acknowledgements}

We thank the anonymous referee for very useful stimulating notes.
We acknowledge V.V. Zhuravlev and G.V. Lipunova for fruitful discussions and 
Max-Planck Institute for Astrophysics (MPA, Garching) for hospitality.   
The work is supported by the Russian Science Foundation grant 14-12-00146.

\bibliographystyle{mn2e}
\expandafter\ifx\csname natexlab\endcsname\relax\def\natexlab#1{#1}\fi
\bibliography{mri}

\appendix
\onecolumn
\section{Linearization of viscous force in dynamical equations}

In cylindrical coordinates for axisymmetric flows the viscous force components read
(see e.g. \cite{1998bhad.conf.....K})
\beq{e:Nr}
{\cal N}_r=\frac{1}{\rho}
\left(
\frac{1}{r}\frac{\partial (rt_{rr}) }{\partial r}
-\frac{t_{\phi\phi}}{r}+\frac{\partial t_{rz}}{\partial z}
\right)\,,
\eeq 
\beq{e:Nphi}
{\cal N}_\phi=\frac{1}{\rho}
\left(
\frac{1}{r^2}\frac{\partial (r^2t_{r\phi}) }{\partial r}
+\frac{\partial t_{z\phi}}{\partial z}
\right)\,,
\eeq 
\beq{e:Nz}
{\cal N}_z=\frac{1}{\rho}
\left(
\frac{1}{r}\frac{\partial (rt_{rz}) }{\partial r}
+\frac{\partial t_{zz}}{\partial z}
\right)\,.
\eeq 

The viscous stress tensor components are:
\beq{e:trr}
t_{rr}=2\eta \frac{\partial u_r }{\partial r}+(\zeta-\frac{2}{3}\eta)\nabla\cdot \bm{u}\,,
\eeq
\beq{e:trphi}
t_{r\phi}=\eta r\frac{\partial (u_\phi/r) }{\partial r}\,,
\eeq
\beq{e:trz}
t_{rz}=\eta \left[
\frac{\partial u_z }{\partial r}+\frac{\partial u_r }{\partial z}
\right]\,,
\eeq
\beq{e:tphiphi}
t_{\phi\phi}=2\eta \frac{u_r}{r}+(\zeta-\frac{2}{3}\eta)\nabla\cdot \bm{u}\,,
\eeq
\beq{e:trphi}
t_{\phi z}=\eta \frac{\partial u_\phi }{\partial z}\,,
\eeq
\beq{e:tzz}
t_{zz}=2\eta \frac{\partial u_z}{\partial z}+(\zeta-\frac{2}{3}\eta)\nabla\cdot \bm{u}\,.
\eeq
(Here $\eta=\rho \nu$ is the dynamical viscosity, $\zeta$ is the second viscosity.).

Below unperturbed and perturbed components will be marked with 
indexes $0$ and $1$, respectively, and therefore
\beq{}
u_r=u_{r,1}, \quad u_\phi=u_{\phi,0}+u_{\phi,1},
\quad u_z=u_{z,1}, \quad \rho=\rho_0+\rho_1, \quad \eta=\eta_0 +\eta_1, \quad T=T_0+T_1\,.
\eeq
For perturbed variables $\rho_1$, $u_{(r,\phi,z),1}$, $T_1$ taken in the form of plane waves $\sim \exp(i\omega t-k_rr-k_zz)$ the partial derivatives simply becomes $\partial/\partial r= -ik_r$, 
$\partial/\partial z= -ik_z$. The dynamic viscosity of interest here is a function of temperature only, $\eta\sim T^{\alpha_{visc}}$, therefore
\beq{}
\frac{\partial \eta}{\partial r}=-\alpha_{visc}\eta_0\frac{T_1}{T_0}ik_r + 
\alpha_{visc}\frac{\eta_0}{T_0}\frac{\partial T_0}{\partial r}\,,
\eeq 
\beq{}
\frac{\partial \eta}{\partial z}=-\alpha_{visc}\eta_0\frac{T_1}{T_0}ik_z + 
\alpha_{visc}\frac{\eta_0}{T_0}\frac{\partial T_0}{\partial z}\,,
\eeq 
(By varying $\eta$, we neglected logarithmic dependence
on temperature and density in the Coulomb logarithm). 

\subsection{Radial component}
Inserting the stress tensor components into \Eq{e:Nr} yields
\beq{}
{\cal N}_r=\frac{1}{\rho}\left[ \eta \frac{\partial^2u_r }{\partial r^2}
+\eta \frac{\partial^2u_r }{\partial z^2}+\frac{\eta}{r}\frac{\partial u_r }{\partial r}-\eta \frac{u_r}{r^2}+2\frac{\partial \eta}{\partial r}\frac{\partial u_r}{\partial r}+\frac{\partial \eta}{\partial z}\frac{\partial u_z}{\partial r}+
\frac{\partial \eta}{\partial z}\frac{\partial u_r}{\partial z} \right]\,.
\eeq
After linearizing we find:
\beq{}
{\cal N}_{r,1}=
\frac{1}{\rho_0}\left[ -\eta_0k^2u_{r,1}-2ik_ru_{r,1}\alpha_{visc} \frac{\eta_0}{T_0}\frac{\partial T_0}{\partial r}-ik_ru_{z,1}\alpha_{visc}\frac{\eta_0}{T_0}\frac{\partial T_0}{\partial z}-ik_zu_{r,1}\alpha_{visc}\frac{\eta_0}{T_0}\frac{\partial T_0}{\partial z}\right]\,.
\eeq
The second term is $\sim k_r/r$ and is small compared to the first term $\sim k^2$ and last two terms $\sim k_z/z$ (we remind that in thin discs considered here $z_0/r\sim u_s/u_{\phi,0}\ll 1$). Noticing that $u_{z,1}k_z+u_{r,1}k_r=0$ from the continuity 
equation, we obtain:
\beq{}
{\cal N}_{r,1}=-\nu k^2u_{r,1}[R]
\eeq
where the factor that takes into account the dependence of viscosity on temperature is defined as
\beq{e:[R]}
[R] =1+i\frac{k_z^2-k_r^2}{k_zk^2}\alpha_{visc}\frac{1}{T_0}\frac{\partial T_0}{\partial z}\,.
\eeq
 
\subsection{Tangential component}
Substituting stress tensor components into \Eq{e:Nphi} yields:
\beq{}
{\cal N}_\phi=\frac{1}{\rho}\left[ \eta \frac{\partial^2u_\phi }{\partial r^2}
+\eta \frac{\partial^2u_\phi }{\partial z^2}+\frac{\eta}{r}\frac{\partial u_\phi }{\partial r}-\eta \frac{u_\phi}{r^2}+\frac{\partial \eta}{\partial r}\frac{\partial u_\phi}{\partial r}-\frac{\partial \eta}{\partial r}\frac{u_\phi}{ r}+
\frac{\partial \eta}{\partial z}\frac{\partial u_\phi}{\partial z} \right]\,.
\eeq
The linearizing leads to:
\begin{eqnarray}
&{\cal N}_{\phi,1}=
\displaystyle\frac{1}{\rho_0}\left[ 
\left(-\alpha_{visc}\eta_0\displaystyle\frac{T_1}{T_0}ik_r+\alpha_{visc}\displaystyle\frac{\eta_0}{r}\displaystyle\frac{T_1}{T_0}\right)\displaystyle\frac{\partial u_{\phi,0}}{\partial r}-\alpha_{visc}\displaystyle\frac{\eta_0}{r^2}\displaystyle\frac{T_1}{T_0}u_{\phi,0}-\eta_0k^2u_{\phi,1}-ik_z\eta_0\displaystyle\frac{1}{T_0}\displaystyle\frac{\partial T_0}{\partial z}u_{\phi,1}\right] \nonumber\\
&-\displaystyle\frac{1}{\rho_0}\displaystyle\frac{\rho_1}{\rho_0}
\left[\left(\alpha_{visc}\displaystyle\frac{\eta_0}{T_0}\displaystyle\frac{\partial T_0}{\partial r}+\displaystyle\frac{\eta_0}{r}\right)\displaystyle\frac{\partial u_{\phi,0}}{\partial r}+
\left(\displaystyle\frac{\alpha_{visc}}{r}\displaystyle\frac{\eta_0}{T_0}\displaystyle\frac{\partial T_0}{\partial r}-
\displaystyle\frac{\eta_0}{r^2}\right)u_{\phi,0}
\right]\,.
\end{eqnarray}
All terms in the second square brackets are $\sim 1/r^2$ and can be neglected
compared to terms in the first square brackets. The latter can be rewritten 
as the sum of two terms:
\beq{}
{\cal N}_{\phi,1}=
\frac{1}{\rho_0}\frac{T_1}{T_0}
\left[ 
\left(
-\alpha_{visc}\eta_0ik_r+\alpha_{visc}\frac{\eta_0}{r}
\right)\frac{\partial u_{\phi,0}}{\partial r}
-\alpha_{visc}\frac{\eta_0}{r^2}u_{\phi,0}
\right]
-\nu k^2 \frac{u_{\phi,1}}{u_{\phi,0}}
\left(1+ik_z\frac{1}{T_0}\frac{\partial T_0}{\partial z}
\right)u_{\phi,0}\,.
\eeq
Here three terms in the first brackets are $\sim k_r/r$, $\sim 1/r^2$, $\sim 1/r^2$, 
respectively, compared to terms $\sim k^2$ in the second brackets, and hence can be neglected. Therefore, we are left with 
\beq{}
{\cal N}_{\phi,1}=-\nu k^2u_{\phi,1}[\Phi]
\eeq
where 
\beq{e:[Phi]}
[\Phi]=1+i\frac{k_z}{k^2}\alpha_{visc}\frac{1}{T_0}\frac{\partial T_0}{\partial z}\,.
\eeq

\subsection{Vertical component}
Substituting the stress tensor components into \Eq{e:Nz} yields:
\beq{}
{\cal N}_z=\frac{1}{\rho}\left[ \eta \frac{\partial^2u_z }{\partial r^2}
+\eta \frac{\partial^2u_z }{\partial z^2}+\left(\frac{\eta}{r}+\frac{\partial \eta}{\partial r}\right)
\left(\frac{\partial u_z }{\partial r}+\frac{\partial u_r }{\partial z}\right)
+2\frac{\partial \eta}{\partial z}\frac{\partial u_z}{\partial z}
\right]\,.
\eeq
The linearization of terms in the middle brackets yields terms $\sim k_r/r, k_z/r$
which are small compared to the term $\sim k^2$ arisen from the second derivatives.
Terms $\sim k_z/z$ arisen from the linearization of the last term, however, should be retained. Therefore, we finally find:
\beq{}
{\cal N}_{z,1}=-\nu k^2u_{z,1}[Z]
\eeq
where 
\beq{e:[Z]}
[Z]=1+2i\frac{k_z}{k^2}\alpha_{visc}\frac{1}{T_0}\frac{\partial T_0}{\partial z}\,.
\eeq

\end{document}